\documentclass[mathleft
]{an}
\usepackage{graphicx}
\usepackage{times}
\begin{document}

\Pagespan{1}{}
\Yearpublication{2012}%
\Yearsubmission{2012}%
\Month{12}%
\Volume{333}%
\Issue{1}%

\newcommand{\ltsima} {$\; \buildrel < \over \sim \;$} 
\newcommand{\simlt}  {\lower.5ex\hbox{\ltsima}}            
\newcommand{\gtsima} {$\; \buildrel > \over \sim \;$} 
\newcommand{\simgt}  {\lower.5ex\hbox{\gtsima}}            
\newcommand{\mdot}{\dot{M}}
\newcommand{\mdotscale}{10^{-8} \ \msun \ {\rm yr^{-1}}}
\newcommand{\msun}{$M_\odot$}
\newcommand{\rsun}{$R_\odot$}
\newcommand{\lsun}{$L_\odot$}
\newcommand{\msyr}{\msun \ {\rm yr^{-1}}}

\def\kms{km~s$^{-1}$}   
\def\e{$\pm$}  
\def\fv{$10^{-14}$}

\title{Simultaneous   $UBVRI$ observations of 
the cataclysmic variable AE Aquarii: temperature and mass of fireballs 
\thanks{based on data collected with 
the telescopes at Bulgarian National Astronomical Observatory Rozhen
and Belogradchick Astronomical  Observatory.}   }

\author{R. K. Zamanov\inst{1}\fnmsep\thanks{Corresponding author:
  \email{rkz@astro.bas.bg}\newline}
\and  G.Y. Latev\inst{1}
\and  K. A. Stoyanov\inst{1}
\and  S. Boeva\inst{1}
\and  B. Spassov\inst{1}
\and  S. V. Tsvetkova\inst{1}
}
\titlerunning{$UBVRI$  observations of fireballs of AE Aqr}
\authorrunning{Zamanov, Latev, Stoyanov, Boeva,  Spassov \& Tsvetkova}
\institute{
     Institute of Astronomy and National Astronomical Observatory, Bulgarian Academy of Sciences,   
       72 Tsarighradsko Shousse Blvd., 1784 Sofia, Bulgaria
      }

\received{5 June 2012}
\accepted{..  August 2012}
\publonline{later}


\abstract{%
  We report simultaneous multicolour observations in  5 bands $(UBVRI)$ 
 of the  flickering variability of  the cataclysmic variable AE Aqr.  
 Our aim is to estimate the parameters (colours, temperature, size) of the fireballs that
 produce the optical flares.  \\
 The observed rise time of the optical flares is in the interval 220 - 440 sec. 
 We estimate  the dereddened colours of the fireballs:
 $(U-B)_0$ in the range 0.8-1.4, $(B-V)_0 \; \sim$~0.03-0.24, $(V-I)_0 \; \sim$~0.26-0.78.  
 We find for the fireballs  a temperature  in the range 10000 - 25000 K, mass  (7-90)$\times 10^{19}$~g,  
 size (3-7)$\times 10^9$~cm  (using a distance of $d=86$~pc).
 These values refer to the peak of the flares observed in $UBVRI$ bands.
  \\
 The data are available upon request from the authors.
 }

\maketitle

\section{Introduction}

AE Aquarii is an 11-12 mag cataclysmic variable, which was
discovered in the optical by Zinner (1938)
and was first associated with the DQ~Her (Intermediate Polar) stars
by Patterson~(1979).  

In  the AE~Aqr system a K0-K4 IV/V star transfers material through the L$_1$ nozzle
toward  a magnetic white dwarf. 
The high-dispersion time-resolved absorption line spectro-  
scopy 
by  Echevarr{\'{\i}}a et al. (2008)
gives the binary parameters as:  
white dwarf mass  $M_{WD} = 0.63 \pm 0.05$ \msun,  
secondary mass $M_2 = 0.37 \pm 0.04 $ \msun, 
binary separation $a = 2.33 \pm 0.02$ R$_\odot$, 
inclination $i \approx 70^o$.

The light curve of AE~Aqr exhibits large flares and coherent oscillations 
of about 16 and 33~s in the optical and X-ray (Patterson 1979). 
It also exhibits radio and millimeter
synchrotron emission (e.g. Bookbinder \& Lamb 1987; Bastian, Dulk \& Chanmugam~1988), 
and possibly even TeV $\gamma$-rays (Bowden et al. 1992; Meintjes et al. 1992). 

It has a long orbital period of 9.88~h (e.g. Casares et al. 1996)
and a very short spin period of the white dwarf.  
On the basis of IUE spectra, Jameson, King \& Sherrington (1980) reveal the most extreme C$_{IV}$/N$_{V}$ ratio 
of all Cataclysmic Variable stars,
probably indicating strong carbon depletion and thus CNO cycling (Mauche, Lee \& Kallman 1997). 
To arrive at such a state,  AE~Aqr is believed to be a 
former supersoft X-ray binary, in which the mass transfer rate in the recent past ($10^7$~yr) has been 
much higher than its current value (Schenker et al. 2002).

\begin{table*} 
\footnotesize
\caption{Journal of observations. In the table are given as follows:
the telescope, band, UT-start and UT-end of the run, exposure time, number of 
CCD images obtained, average magnitude in the corresponding band, 
minimum -- maximum magnitudes in each band, standard
deviation of the mean, typical observational error.}
\begin{center}
\begin{tabular}{llrrrrcrrccccrlcr}
\hline

Telescope& Band       &   UT        & Exp-time  &  N$_{pts}$  &  average & min-max     & stdev  & err   \\
      &               & start-end   & [sec]     &             & [mag]    & [mag]-[mag] & [mag]  & [mag] \\ 
\hline
{\bf 2010 Aug 12 }  &     & JD2455421  \\

 50/70 cm Schmidt    & $B$ & 23:20 - 00:32 & 60,90,120 & 42 & 12.338 & 12.214 - 12.397 & 0.047 & 0.010 \\
60 cm Belogradchick  & $R$ & 23:00 - 00:37 & 40        & 81 & 10.717 & 10.607 - 10.754 & 0.027 & 0.007 \\
\hline  
{\bf 2010 Aug 13 } &     & JD2455422  \\ 
 2.0 m RCC         & $U$ & 22:23 - 00:28 & 180       & 19 & 11.577 & 10.985 - 12.335 & 0.336 & 0.025 \\
 50/70 cm Schmidt  & $B$ & 21:46 - 00:28 &  60       &137 & 12.226 & 11.694 - 12.430 & 0.169 & 0.008 \\
 2.0 m RCC         & $V$ & 22:23 - 00:28 &  10       &463 & 11.377 & 10.990 - 11.486 & 0.107 & 0.007 \\
 60 cm Belogradchick    & $R$ & 22:00 - 00:28 & 60,40     &181 & 10.699 & 10.422 - 10.797 & 0.077 & 0.006 \\
 60 cm Rozhen      & $I$ & 22:41 - 00:29 & 15        &388 & 10.177 &  9.987 - 10.261 & 0.052 & 0.007 \\
\hline
{\bf 2010 Aug 14 } &     & JD2455423  \\
 2.0 m RCC         & $U$ &  18:57 - 23:43 & 120,180    & 67   & 12.073 & 10.824 - 12.942 & 0.527 & 0.022 & \\
 50/70 cm Schmidt  & $B$ &  18:44 - 23:44 & 60,120,180 & 245  & 12.186 & 11.672 - 12.450 & 0.144 & 0.011 & \\
 2.0 m RCC         & $V$ &  18:48 - 23:45 & 10         & 1106 & 11.346 & 11.056 - 11.831 & 0.080 & 0.007 & \\
 60 cm Belogradchick      & $R$ &  19:32 - 23:46 & 40,60      & 331  & 10.639 & 10.432 - 10.749 & 0.057 & 0.005 & \\
 60 cm  Rozhen     & $I$ &  18:24 - 23:42 & 15         & 1132 & 10.143 & 9.976 - 10.284  & 0.060 & 0.009 & \\
\hline
{\bf 2010 Aug 16 } &     & JD2455425  \\
60 cm  Rozhen      & $U$ &  18:31 - 19:59 & 120        & 24   & 12.510 & 12.013 - 13.100 & 0.278 & 0.085 & \\
60 cm  Rozhen      & $B$ &  18:33 - 00:26 & 60,90      & 76   & 12.317 & 12.016 - 12.565 & 0.142 & 0.012 & \\
60 cm  Belogradchick       & $V$ &  18:54 - 00:27 & 60,180	 & 68	& 11.383 & 11.207 - 11.563 & 0.119 & 0.008 & \\
60 cm  Belogradchick       & $R$ &  18:57 - 00:28 & 40	 & 71	& 10.738 & 10.565 - 10.917 & 0.109 & 0.006 & \\
60 cm  Belogradchick       & $I$ &  18:53 - 00:29 & 40	 & 70	& 10.190 & 10.036 - 10.351 & 0.095 & 0.006 & \\
\hline
{\bf 2011 Aug 31 } &     & JD2455805 \\
 2.0 m RCC          & $U$ &  20:59 - 23:46 & 90,120 & 72   & 11.368 &  9.841 - 12.476 & 0.685 & 0.039 & \\
 50/70 cm Schmidt   & $B$ &  20:54 - 23:40 & 30     & 270  & 11.777 & 10.767 - 12.339 & 0.376 & 0.032 & \\
 2.0 m RCC          & $V$ &  20:58 - 23:18 & 5      & 1090 & 11.079 & 10.355 - 11.738 & 0.228 & 0.012 & \\
 60 cm  Rozhen      & $R$ &  20:45 - 23:44 & 10     & 328  & 10.421 &  9.876 - 10.730 & 0.175 & 0.015 & \\
 60 cm  Rozhen      & $I$ &  20:45 - 23:44 & 10     & 327  &  9.950 &  9.546 - 10.195  & 0.129 & 0.015 & \\
\hline

\end{tabular}
\end{center}
\label{tab.journal}
\end{table*}

The Balmer emission lines vary both in
strength and shape and they are not good tracers of the orbital
motion of the white dwarf. This has led to the proposal of the
magnetic propeller model (Wynn, King \& Horne 1997). 
The gas stream emerging from the K2V companion star through the 
inner Lagrangian point L1 encounters a rapidly spinning magnetosphere. 
The rotating white dwarf in AE~Aqr ejects most of the matter transferred
from the secondary in the form of  blobs  
(\textquoteleft  fireballs\textquoteright ).
Only a small fraction ($\sim$3\%) of the mass flow at the magnetospheric radius eventually accretes 
on to the surface of the white dwarf, emphasizing the  effective magnetospheric propeller process in the system
(Oruru \& Meintjes 2012).

The white dwarf in AE Aqr is a fast rotator  having a spin period 
$P=33.08$~s, and  spinning down at a rate $5.64 \times 10^{-14}$~s~s$^{-1}$ 
(de Jager et al. 1994; Mauche et al. 2011), corresponding to a spin-down luminosity of 
$6 \times 10^{33}$~erg~s$^{-1}$  (Oruru \& Meintjes 2011). 
A part of the spin-down power is consumed to expel the blobs.

In this paper we present simultaneous multicolour observations in  5 bands $(UBVRI)$ of the flares of AE~Aqr.
The source of the flares is assumed to be the fireballs(blobs). 
We measure the fluxes emitted by the fireballs, their colours, and  calculate
their mass, size, temperature and expansion rate.

\section{Observations} 

We observed AE~Aqr with 4 telescopes mixing from 2010 August to 2011 August
-- see Table~\ref{tab.journal} for details. 

All telescopes were  equipped with CCD cameras: 
the 2m RCC telescope of the National Astronomical Observatory  Rozhen 
was equipped with a dual channel focal reducer --  
in the $U$ band a Photometrics CE200A CCD (1024x1024 px, field of view 7.5'x7.5')
has been used, and in the $V$ band a VersArray 1330B (512x512 px, 7.5'x7.5');   
the 
50/70 cm Schmidt telescope was equipped with FLI PL16803 CCD, 4096x4096 px, used 1024 x1024 px, 18'x18';
the 60 cm Rozhen telescope  was equipped with a FLI PL09000 CCD with 3056x3056 px and 18'x18' and
the 60 cm  Belogradchick telescope was equipped with FLI PL 09000 CCD, 3056x3056 px, 18'x18'.

\begin{figure*}
 \vspace{21.0cm}     
    \includegraphics{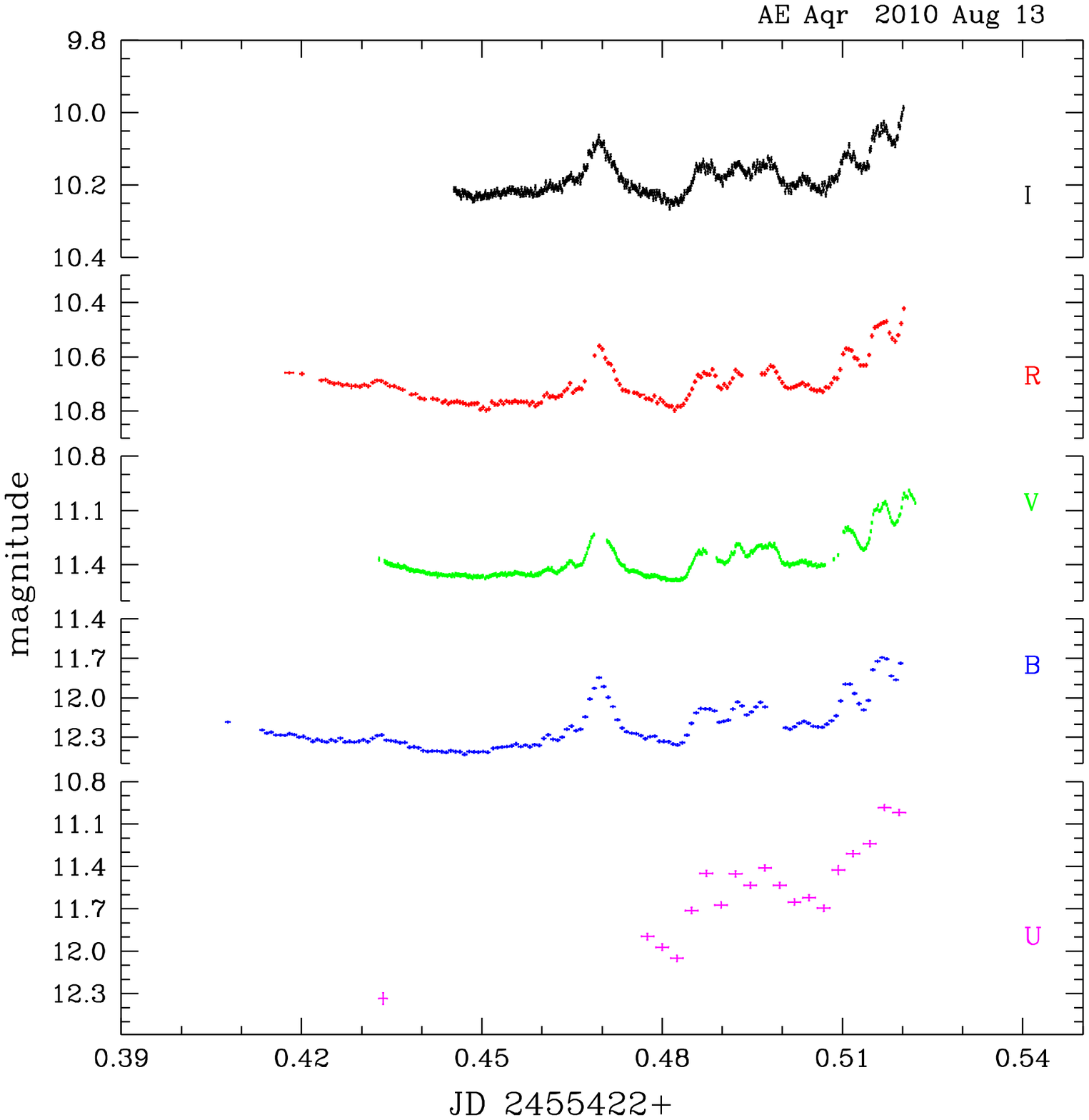}   
    \includegraphics{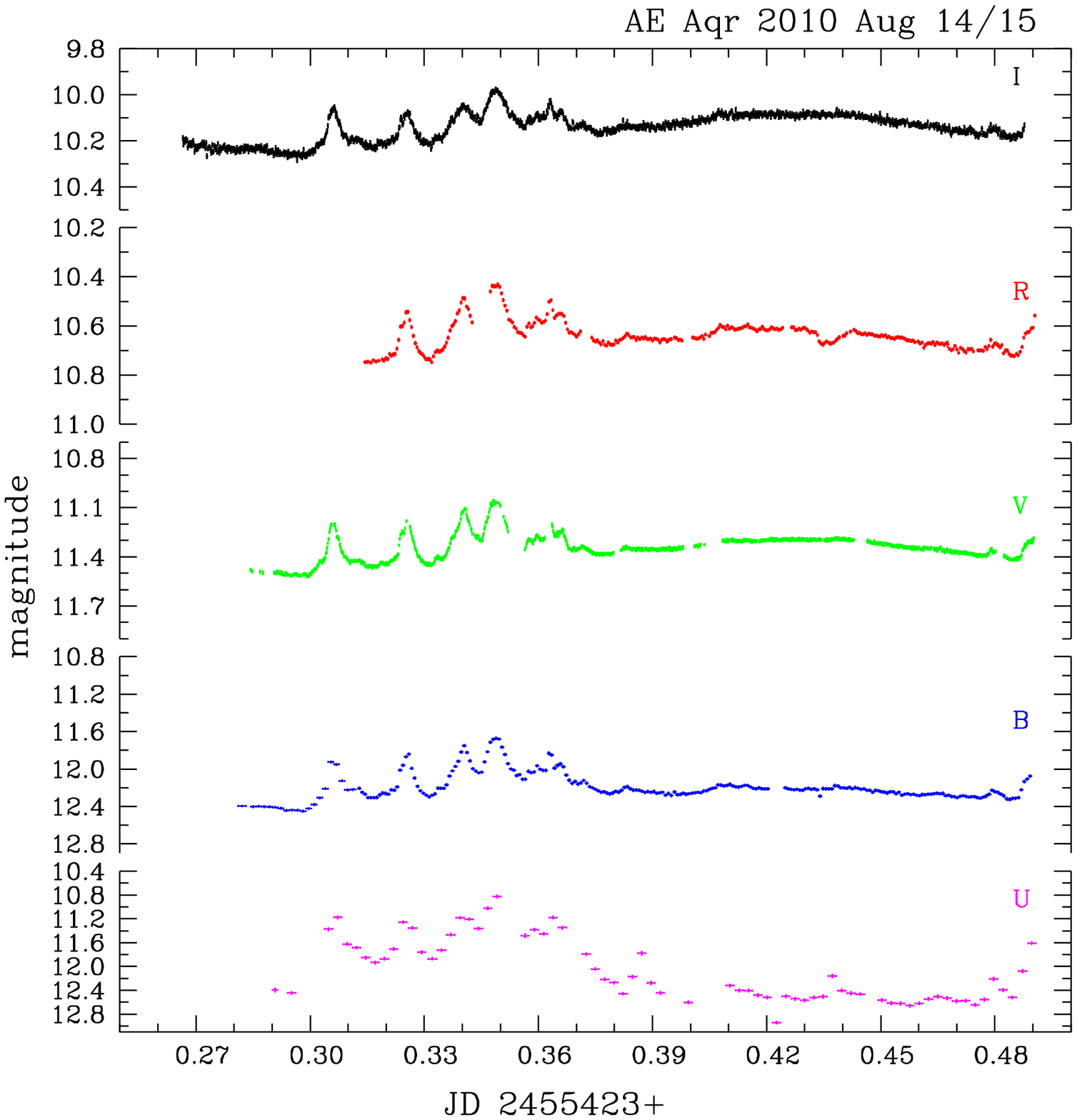}   
    \includegraphics{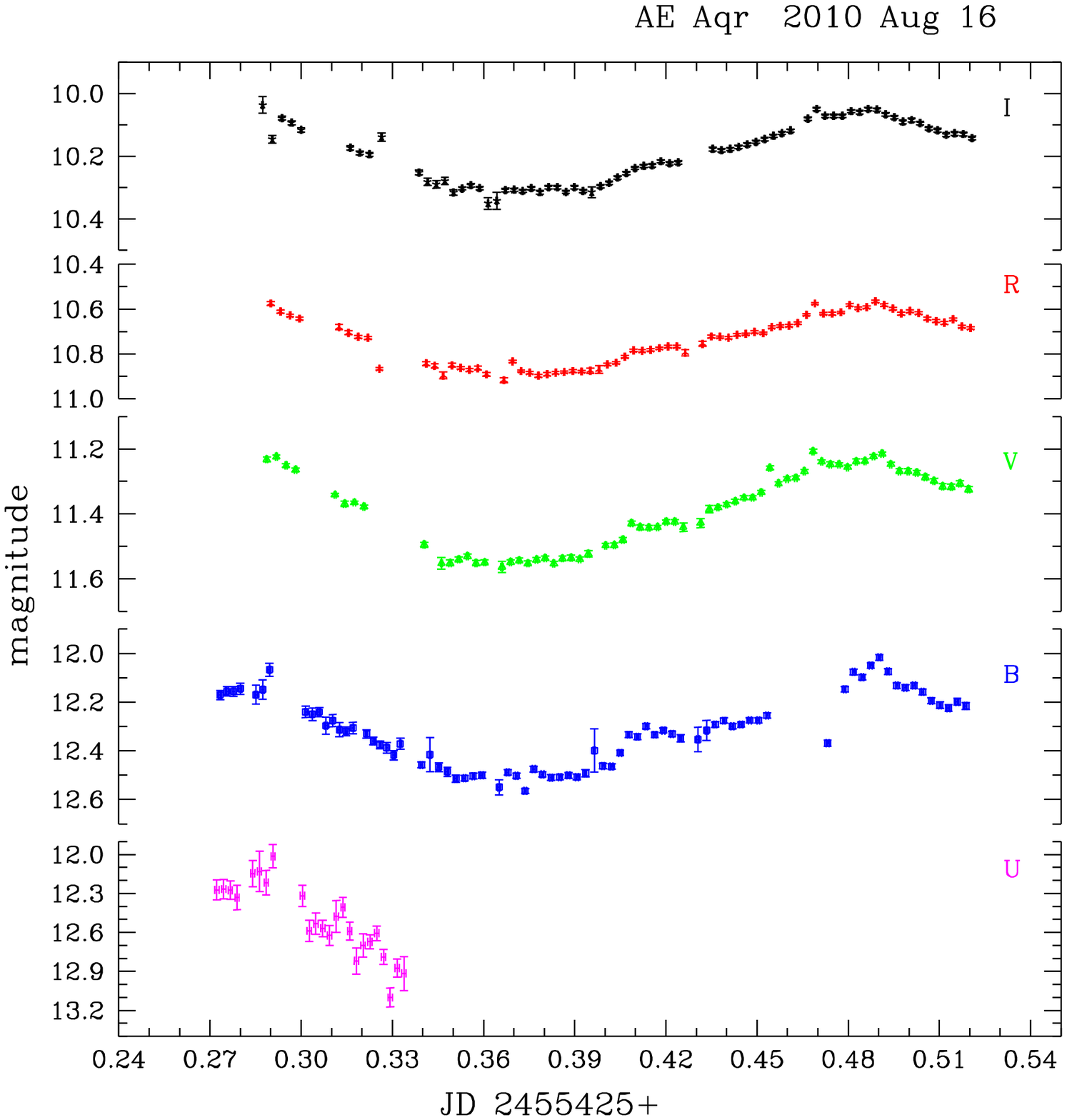}   
    \includegraphics{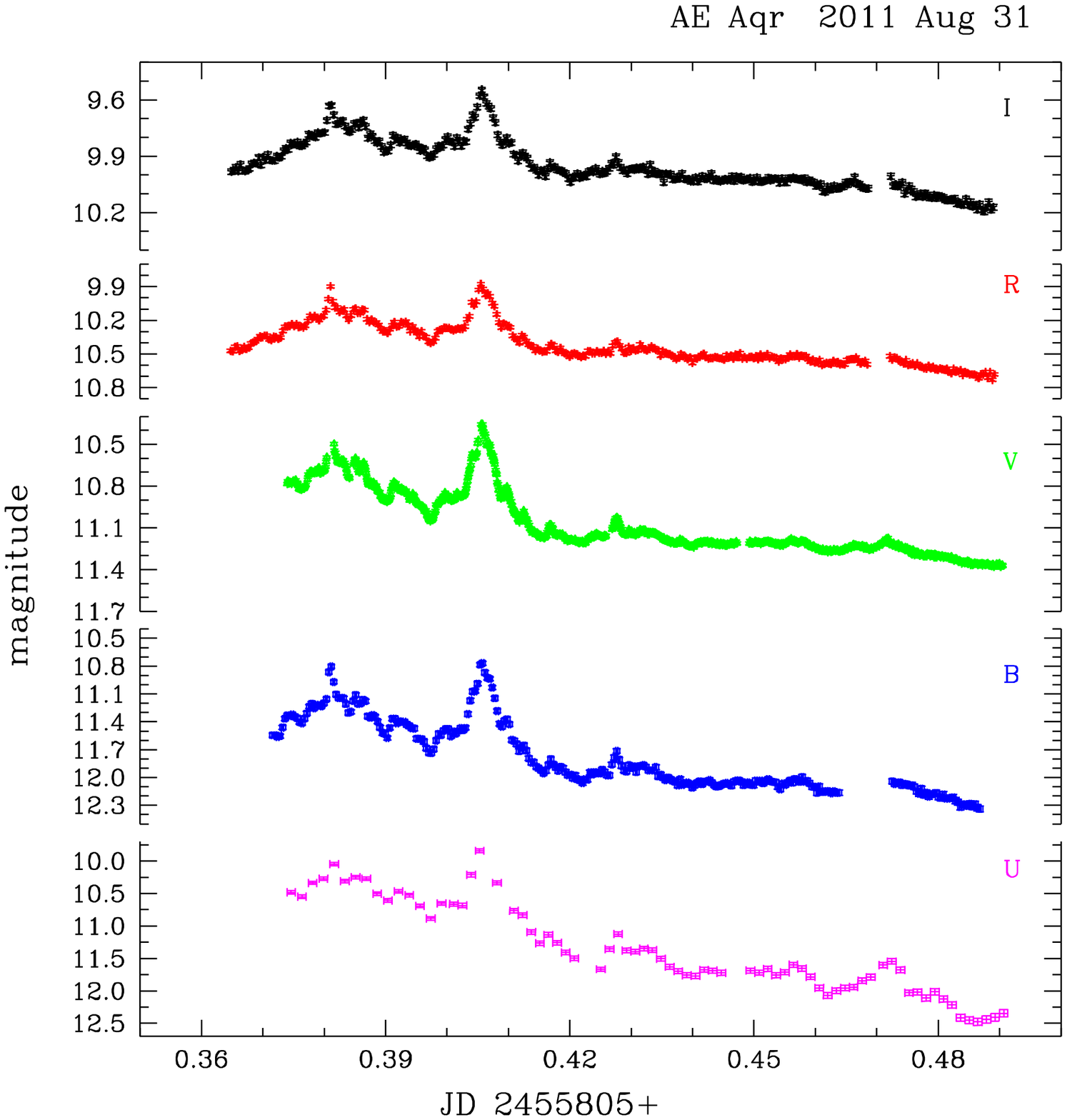}  
 \caption[]{Variability of AE~Aqr in the $UBVRI$ bands 
     on 2010 August 12, 
        2010 August 13, 
        2010 August 14, 
   and  2011 August 31. 
  }		    
\label{fig.UBVRI}     
\end{figure*}

The journal of observations is given in Table~\ref{tab.journal}. 
For each run  we also give 
\hskip 0.1cm 
 the minimum, maximum, and  
average  brightness in  the corresponding band, plus the standard deviation of the run. 
All the CCD images have been bias subtracted, flat fielded, and standard 
aperture photometry has been performed. The data reduction and aperture photometry 
are performed with IRAF\footnote{ IRAF is distributed by
  the National Optical Astronomy Observatory, which is operated by the
  Association for Research in Astronomy, Inc.  under cooperative
  agreement with the National Science Foundation.} and have been checked with alternative software packages. 
 The light curves are plotted in Fig.\ref{fig.UBVRI}.

\begin{figure*} 
 \vspace{10.8cm}  
  \includegraphics{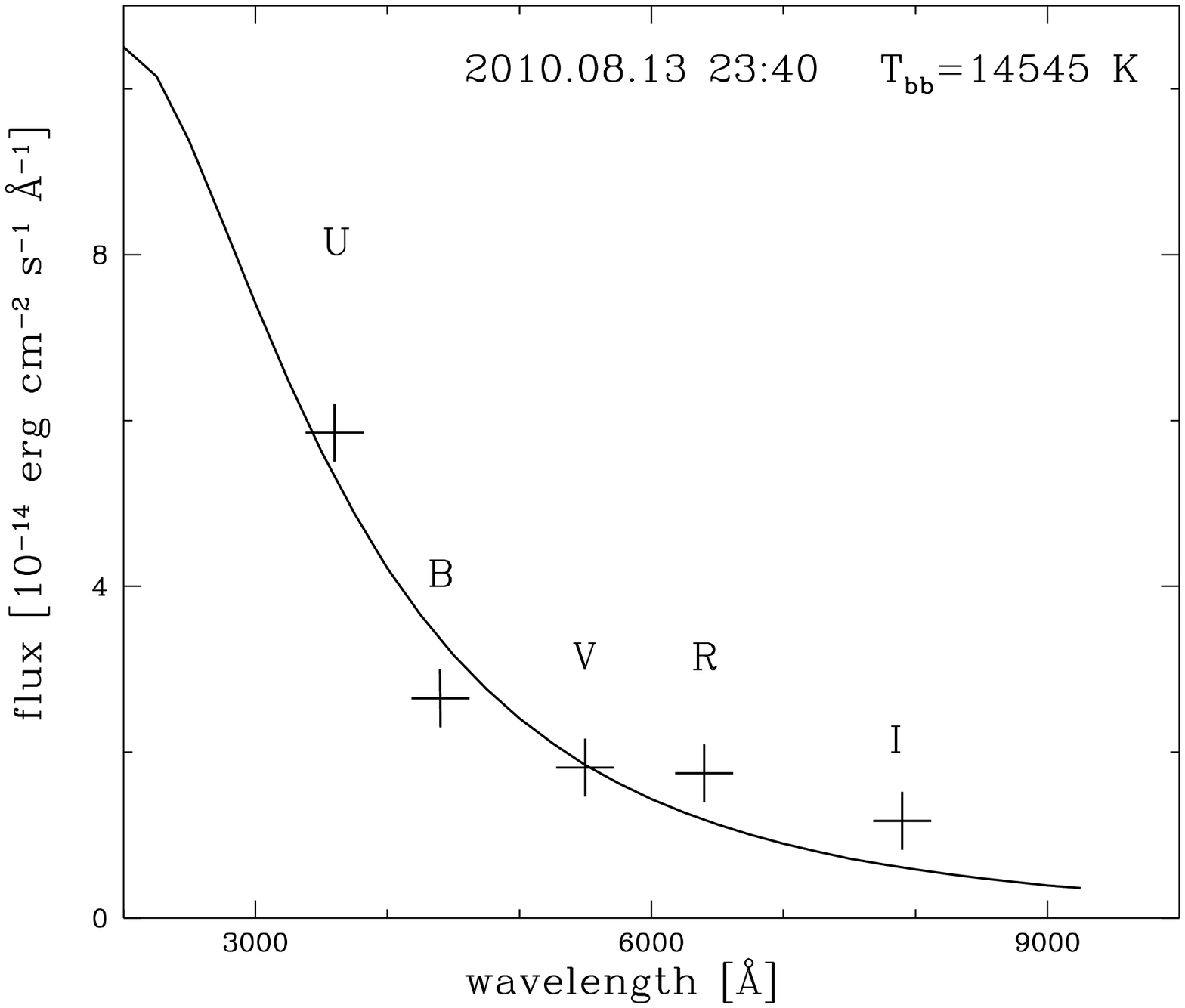}   
  \includegraphics{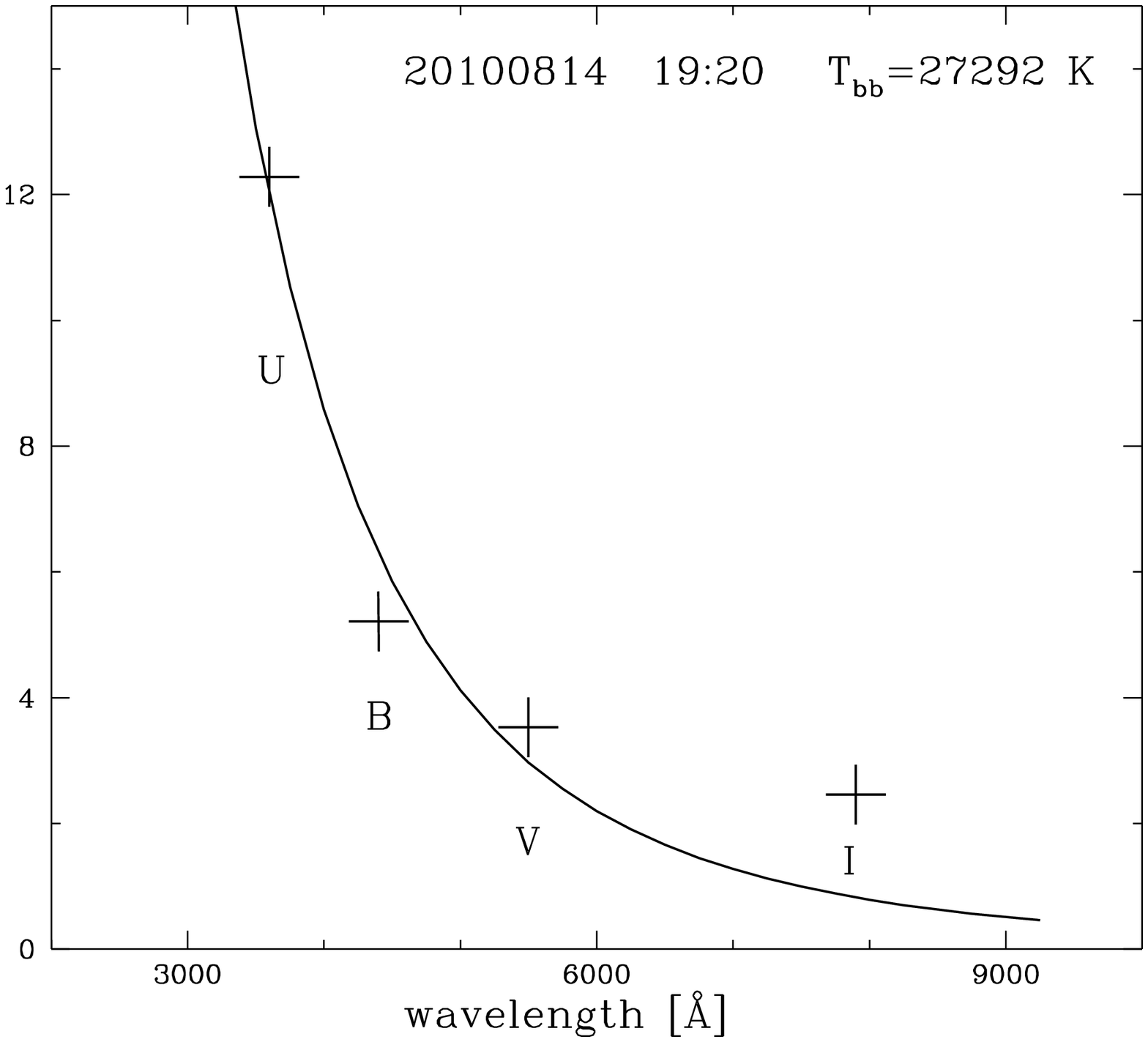}   
  \includegraphics{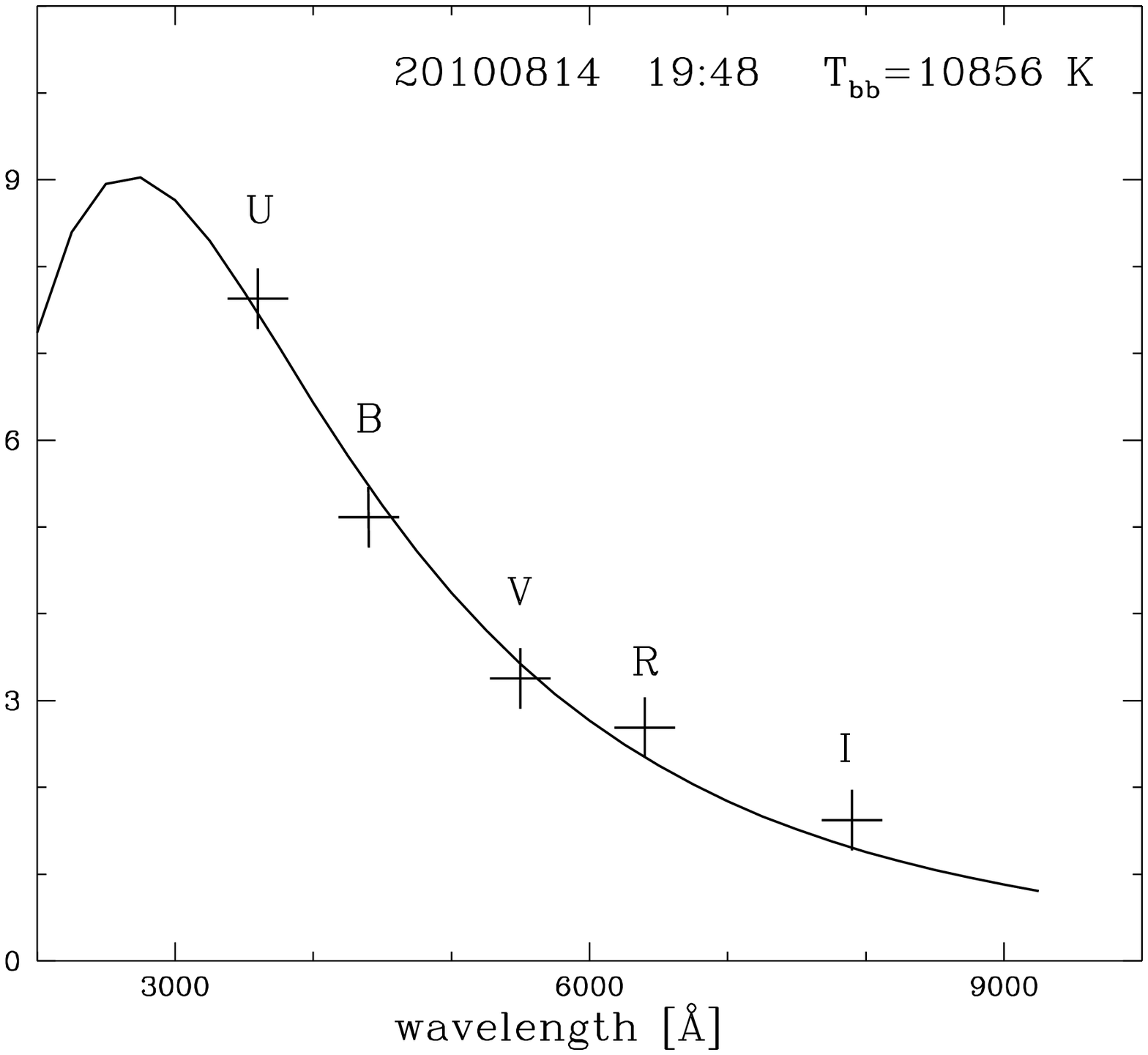} 
  \includegraphics{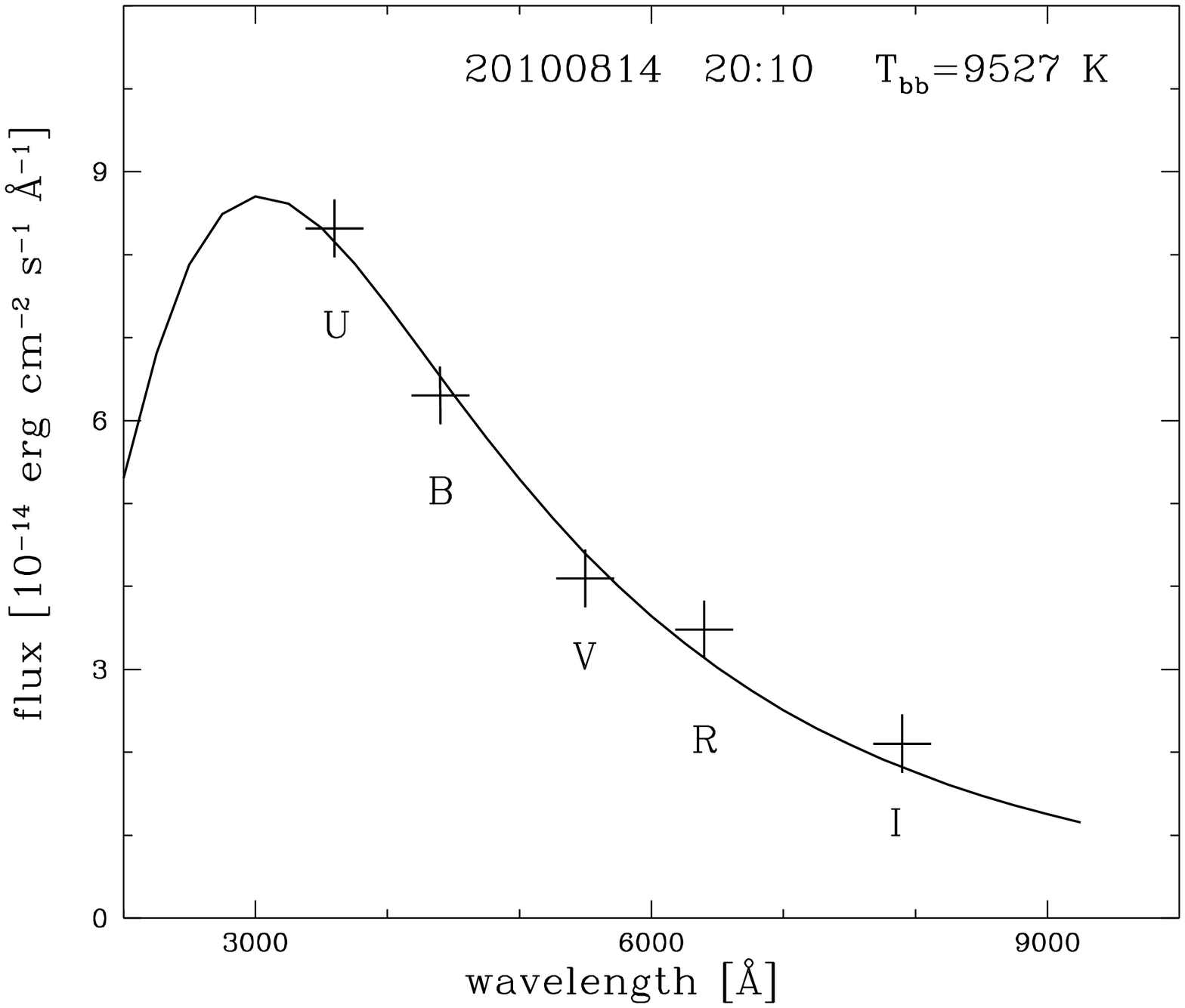} 
  \includegraphics{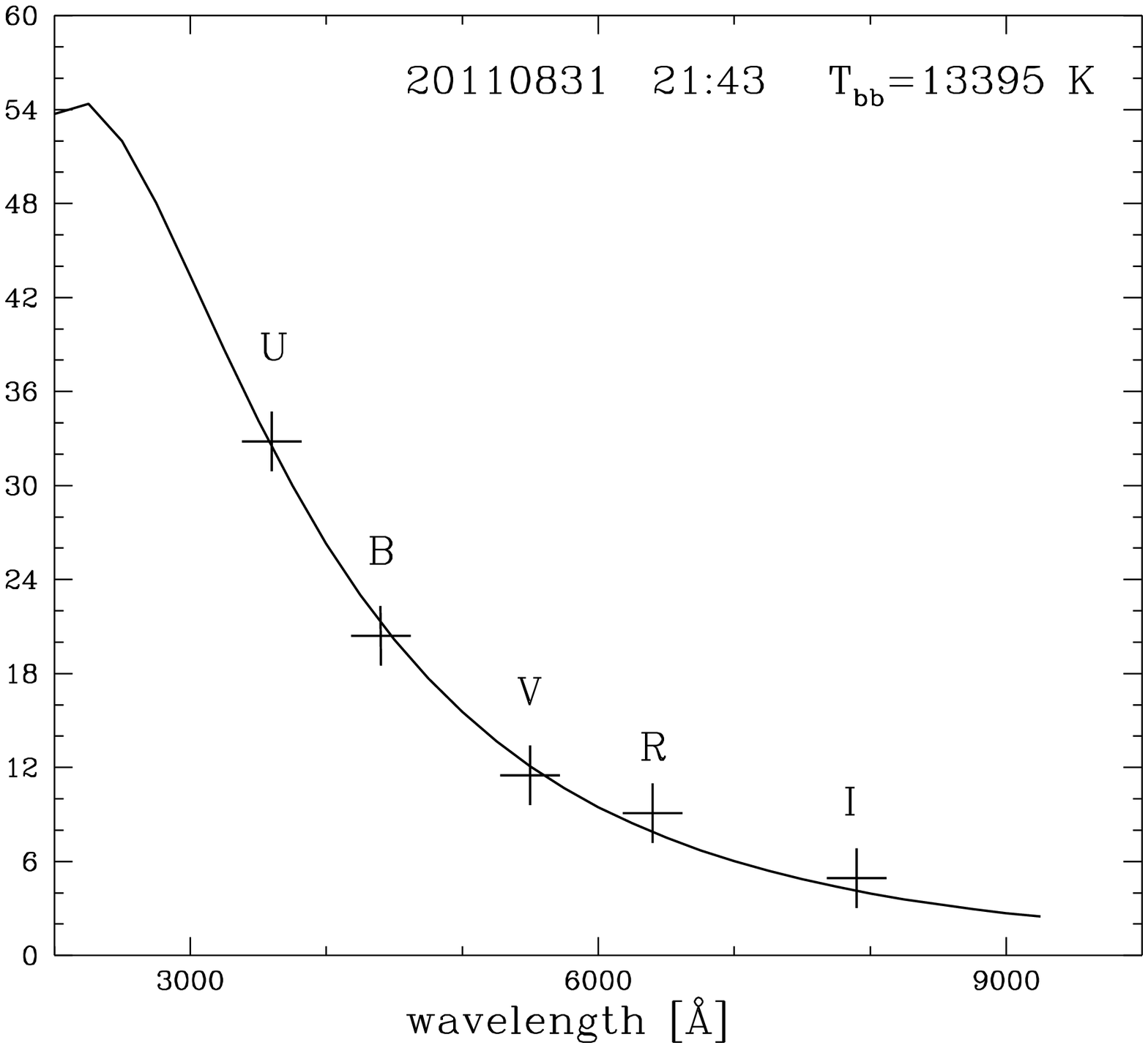} 
  \label{fig.flux}
\caption{Dereddened fluxes of the fireballs at the peak of the flare. The solid line 
    represents a black body fit. }
 \vspace{11.5cm}   
    \includegraphics{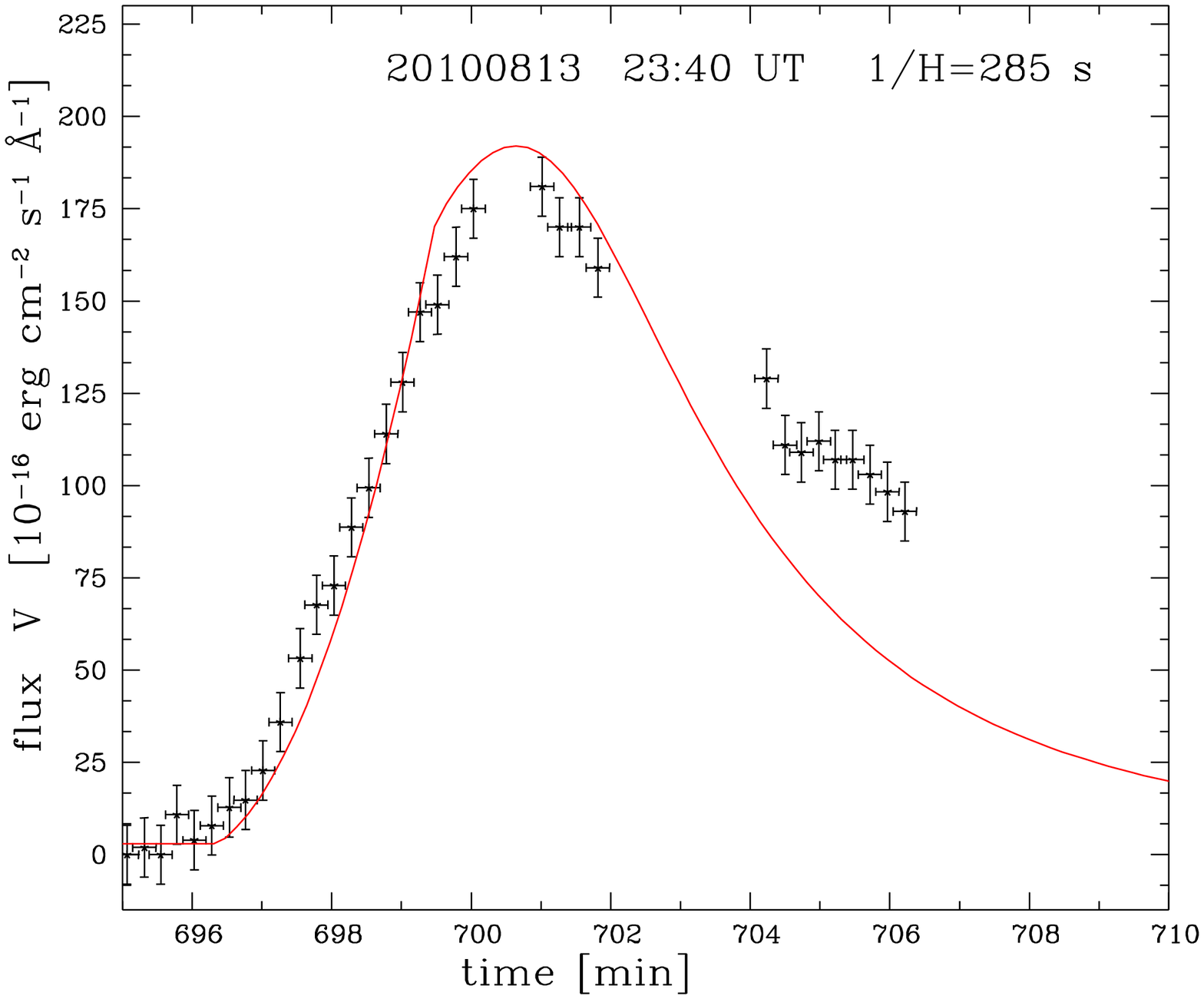}   
    \includegraphics{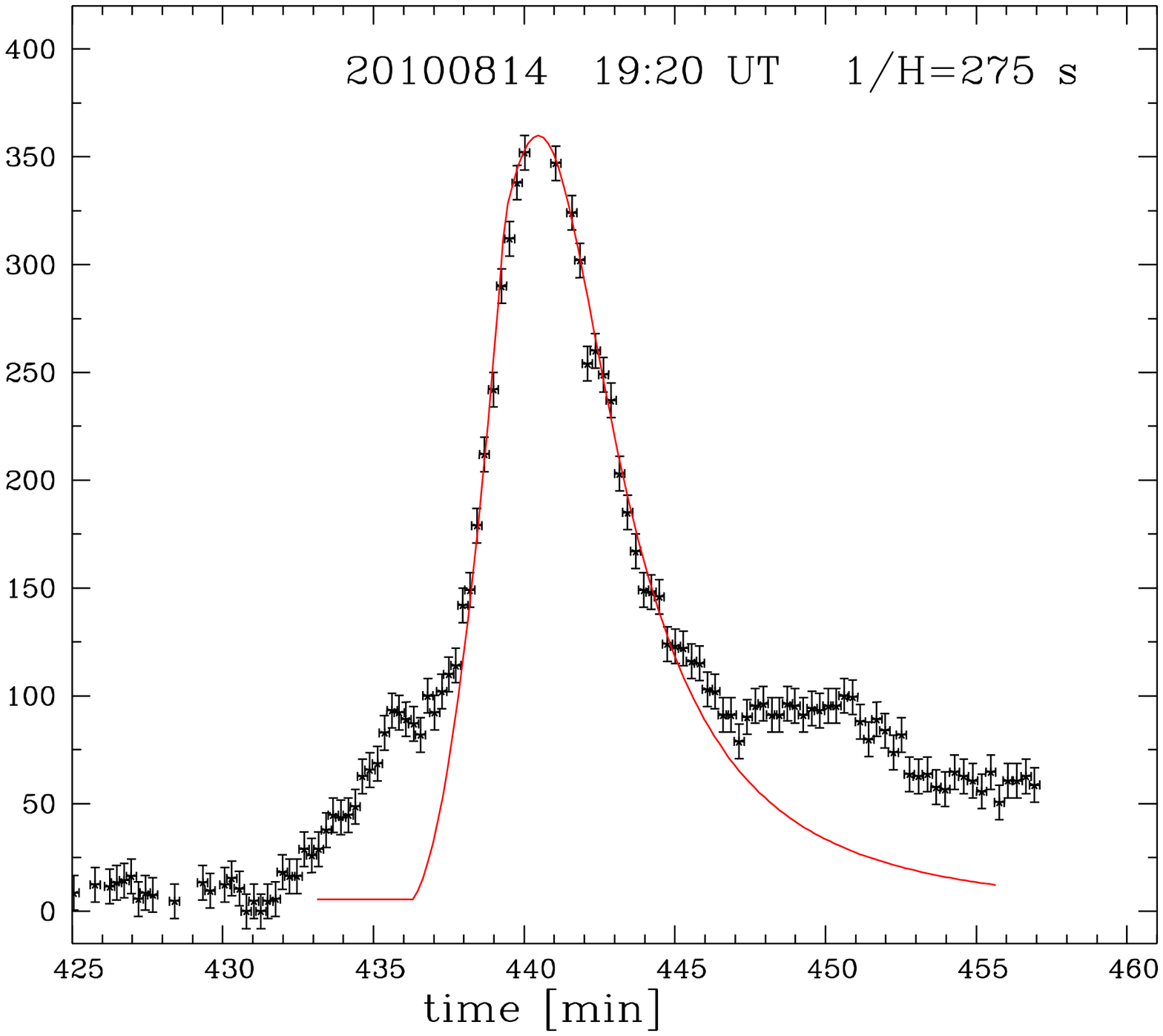}   
    \includegraphics{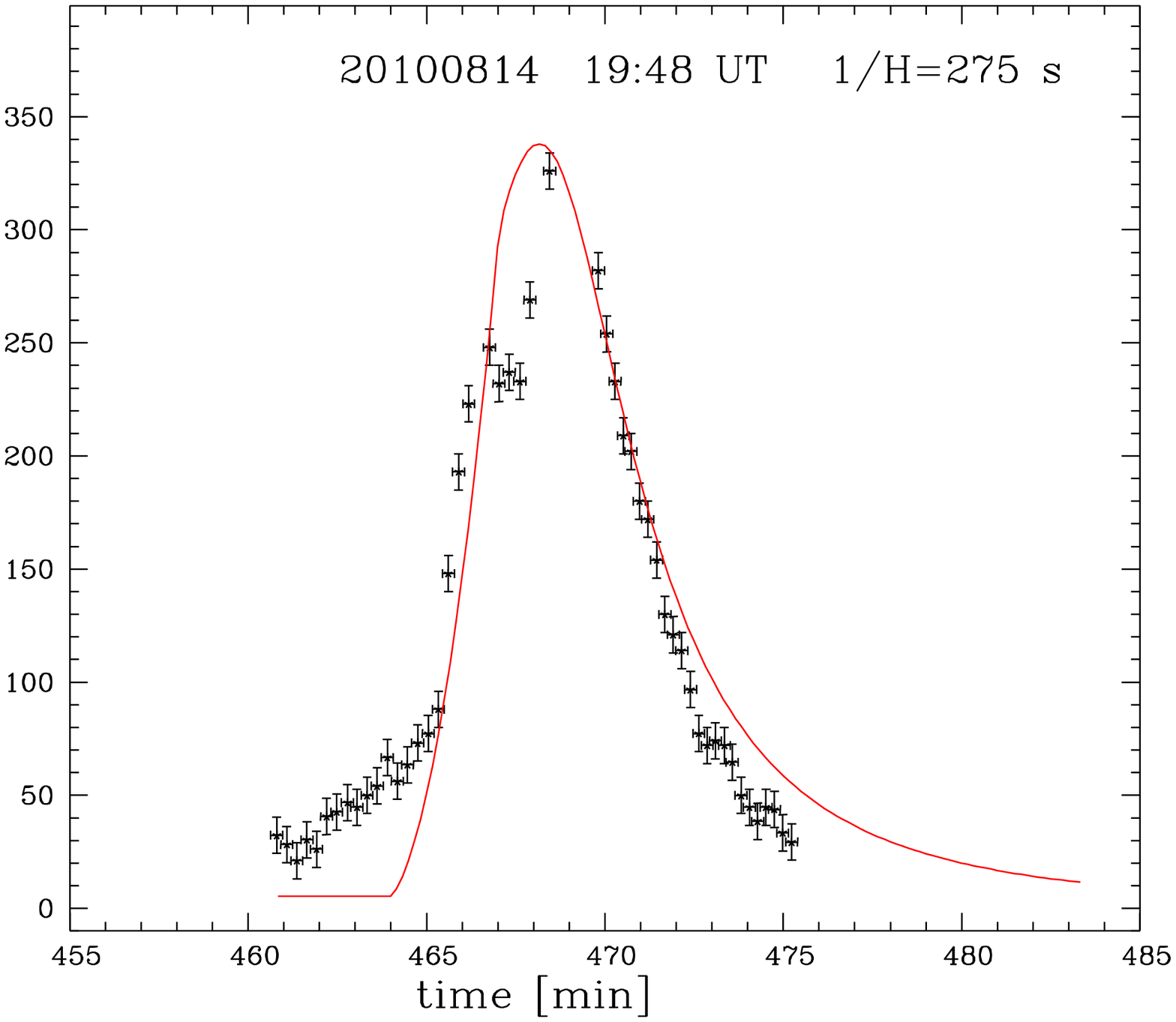} 
    \includegraphics{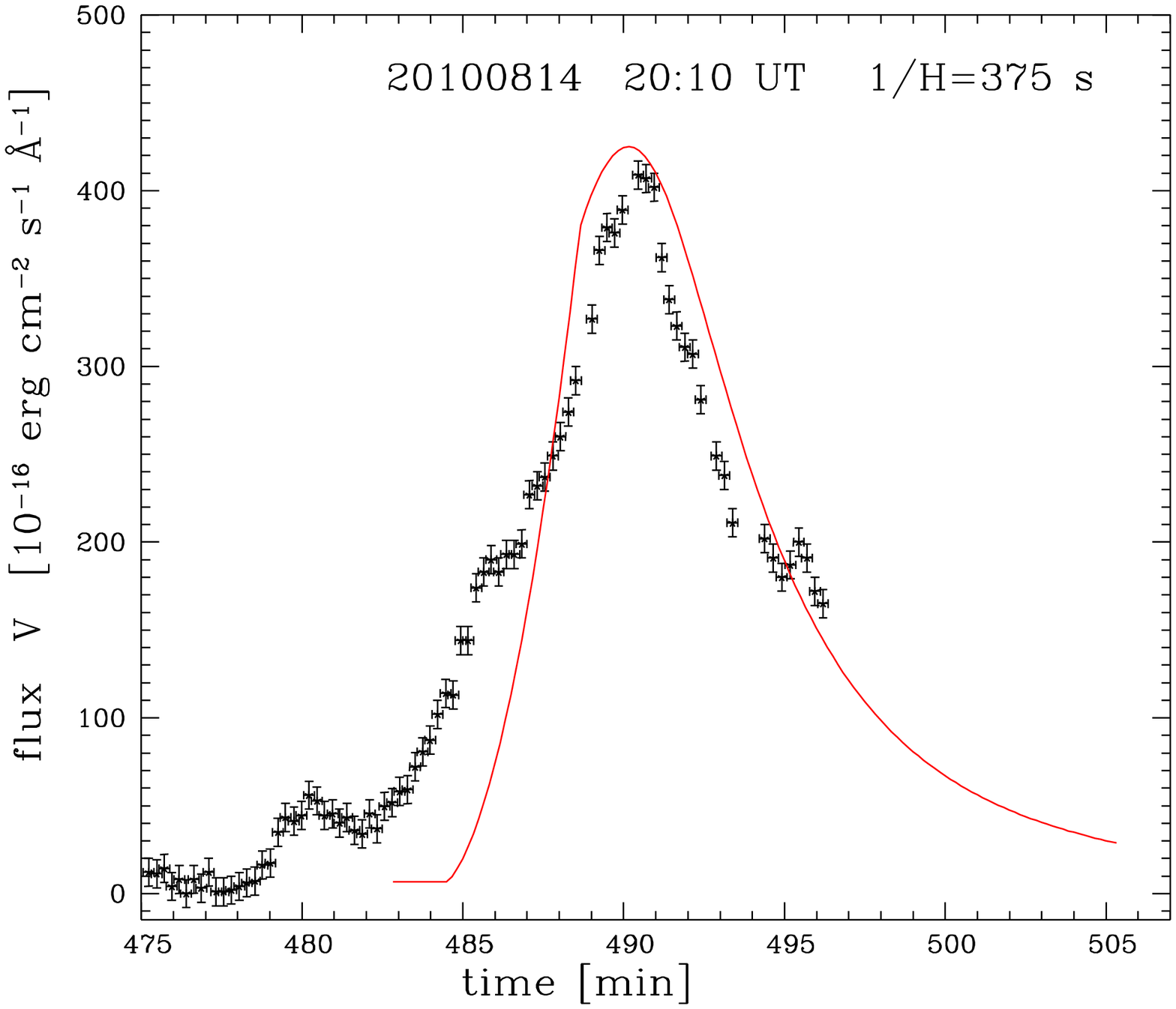} 
    \includegraphics{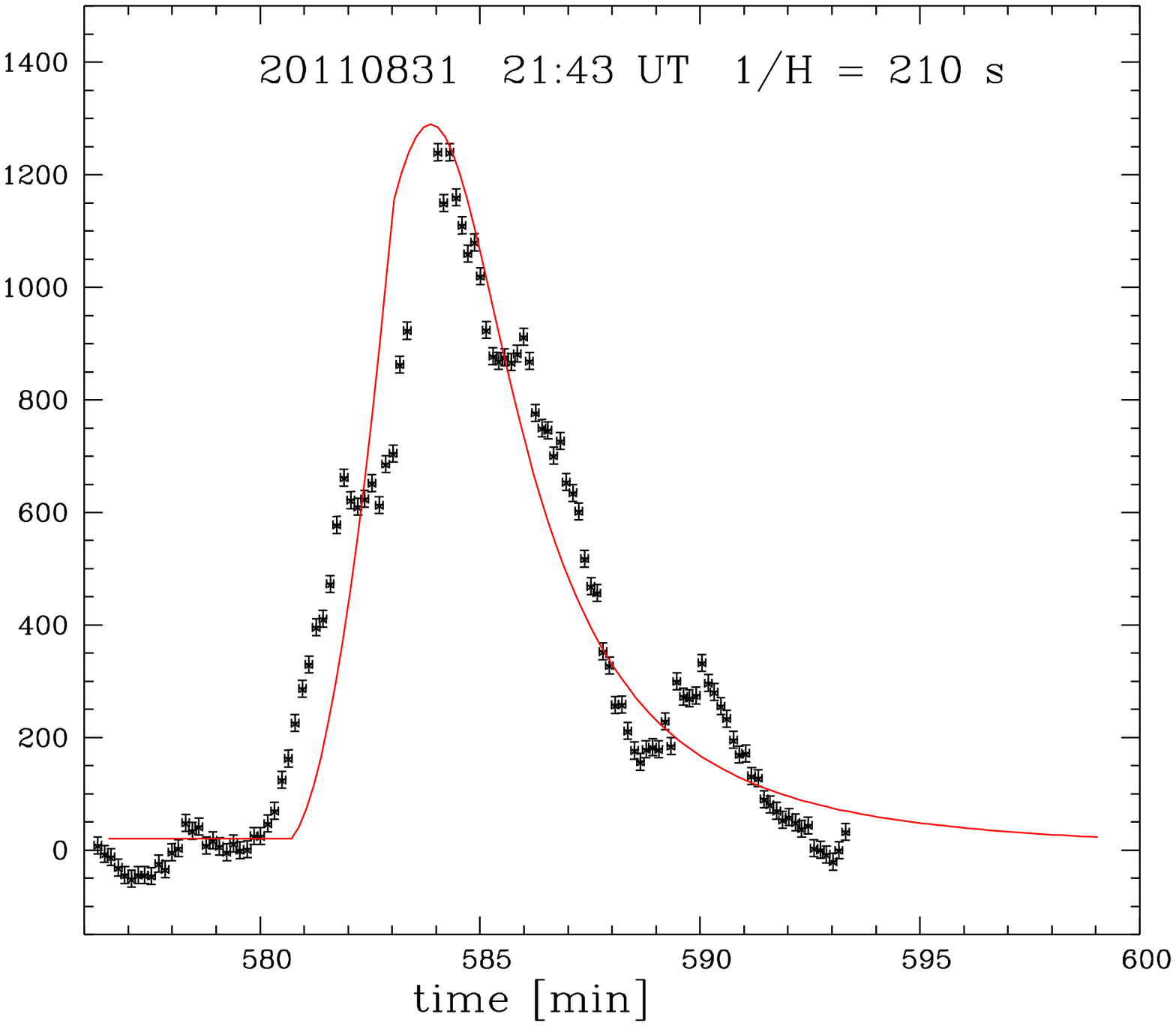} 
\caption{Model of the time evolution of the V band flux.  The solid line is the model 
  of isothermal fireball with the corresponding parameters as listed in Table~\ref{Tab.res}.}
\label{fig.models}
\end{figure*}

\begin{table*}
\centering
\caption{The calculated parameters of the individual fireballs. In the table are given
as follows: rise time in seconds, dereddened peak flux of the fireball in  $UBVRI$  bands, 
dereddened colours of the peak emission of the fireball, 
calculated temperature of the fireball, its size, mass, expansion  velocity, expansion constant,
central density (at the peak of the flare).  }
\begin{tabular}{ccccccccrlr}
  \hline

          &   20100813 23:40  &  20100814 19:20  &  20100814 19:48  &   20100814 20:10 &   20110831 21:43 &  \\
Quantity  &   JD 2455422.486  &  JD  2455423.305    &   JD  2455423.325   &   JD  2455423.340   &  JD 2455805.405     &  \\

\hline
\\
rise time [sec]      & 260\e20  &   230\e25  & 290\e30   &  440\e20  &   260\e30  \\
\\
F$_U$  [\fv\ erg cm$^{-2}$ s$^{-1}$ \AA$^{-1}$]  & 5.85 & 12.28 & 7.63 & 8.32 & 32.8        \\
F$_B$  [\fv\ erg cm$^{-2}$ s$^{-1}$ \AA$^{-1}$]  & 2.65 &  5.21 & 5.11 & 6.30 & 20.4        \\
F$_V$  [\fv\ erg cm$^{-2}$ s$^{-1}$ \AA$^{-1}$]  & 1.81 &  3.53 & 3.26 & 4.09 & 11.5        \\
F$_R$  [\fv\ erg cm$^{-2}$ s$^{-1}$ \AA$^{-1}$]  & 1.75 &  ---  & 2.69 & 3.48 & 9.10        \\
F$_I$  [\fv\ erg cm$^{-2}$ s$^{-1}$ \AA$^{-1}$]  & 1.17 &  2.46 & 1.62 & 2.10 & 4.93        \\

\\

$(U-B)_0$            &  -1.36\e0.06 & -1.43\e0.03 & -0.93\e0.04 & -0.80\e0.05 & -1.02\e0.07 \\
$(B-V)_0$            &   0.24\e0.03 &  0.23\e0.03 &  0.17\e0.03 &  0.19\e0.03 &  0.03\e0.06 \\ 
$(V-R)_0$            &   0.47\e0.05 &  ---        &  0.30\e0.03 &  0.33\e0.03 &  0.26\e0.03 \\ 
$(V-I)_0$            &   0.70\e0.07 &  0.78\e0.03 &  0.42\e0.06 &  0.45\e0.05 &  0.26\e0.06 \\
\\

\hline 

\\

temperature  T  [K]        &  14 545\e1000   &    27 292\e1500   &   10 856\e150  &    9 527\e100 &     13 395\e200   \\
size $a_{pk}$ [$10^9$ cm]  &    3.0\e0.3     &      2.5\e0.3     &    5.3\e0.3    &      7.1\e0.4 &       7.7\e0.4    \\
\\
mass $M$ [$10^{19}$ g]     &    9.6\e1.5    &       6.8\e1.5     &    39 \e 6     &     78\e12    &       97\e15      \\ 

\\

expansion velocity  $v$ [\kms ]  & 105\e10  &  91\e10  &  193 \e 18  & 189\e18   & 367\e 30  \\
expansion constant  1/H [s]      & 285\e20  & 275\e20  & 275\e20  & 375\e25  &  210\e20      \\

\\

central density $\rho$ [$10^{-10}$ g cm$^{-3}$] & 6.4   & 7.8 &   4.7 &  3.9  &  3.8  \\

\\
\hline
 
\end{tabular}
 \label{Tab.res}
\end{table*}

\section{Analysis}

\subsection{Flares of AE~Aqr}

The light curve of AE~Aqr exhibits large flares on time-scales of about 10 minutes, 
which were first reported by Henize (1949).  
The first simultaneous multicolour optical photometry  was performed 
by Chincarini \& Walker (1981). 
Later  van Paradijs, van Amerongen,  \& Kraakman (1989), performed 
five-colour  (Walraven system) observation  and demonstrated
that the flares occur throughout the whole orbital period and estimated the rise time $\sim100-200$~s. 
Bruch \& Grutter (1997) found that the probability for strong flares is phase dependent.
Mauche et al. (2011) performed a multiwavelength (TeV $\gamma$-ray, X-ray, UV, optical, and radio)
campaign of observations of AE~Aqr. They demonstrated that there is a strong correlation in 
the flux variations between the X-ray, near UV, and optical B and V bands. They also detected a weak 
negative correlation between X-ray and radio wavebands.

The observed properties of the close binary AE Aqr indicate that the mass transfer in this system 
operates via the Roche lobe overflow mechanism, 
\hskip 0.3mm 
but the material transferred from the normal companion 
is neither accreted onto the surface of the white dwarf nor stored in a disk around its magnetosphere. 
In terms of our current understanding (Wynn, King \& Horne 1997 ), 
the short time-scale variations or flares can best be explained in terms
of a fragmented accretion flow. The accretion from the companion star is fragmented into discrete blobs 
that interact with the  propeller that is the
magnetosphere of the  spinning white dwarf. 
These blobs (fireballs) are the reason for the flares. 
A typical flow pattern is: blobs reach a maximum velocity of \simlt $1000$ \kms \
at closest approach to the white dwarf (\simgt 10$^{10}$ cm), and cruise out of the system
with a velocity of $V \approx 300$~\kms .
The agreement between the observed H$\alpha$  Doppler tomogram of AE~Aqr and the predicted one  
suggests that the blobs  remain intact until they pass closest approach and
are ejected (Wynn, King \& Horne 1997).

\subsection{Flux of the flare}

Our aim is to measure the parameters (temperature, mass, radius) of the blobs (fireballs).
To do this we need to measure the magnitude at the start and at the peak of the flare. 

During our first night of observations (2010 Aug 12) we do not detect any flares. 
The same is true during the fifth night (2010 Aug 16). Only smooth variations due to the orbital modulation are visible.
More details  about the orbital modulation are given in van Paradijs et al. (1989). 

During the second night (2010 August 13) we detect 7 flares (see Fig.\ref{fig.UBVRI}). Among them 
one has well defined start and peak in  $UBVRI$  bands. During the
night of 2010 August 14  there are 5 flares;  we  are able to measure 3 of them. 
On the night  2011 August 31 we have one well defined flare. 
The rise time of these five flares is given in  Table~\ref{Tab.res}. 
The rise time is measured as the time from the beginning of the flare to its peak.

For a few other flares, that are not listed in Table~\ref{Tab.res}, we measure the rise time as follows:
flare at 0813~23:15  rise time  $265  \pm 20$~s,
flare at 0813~0:14  $220 \pm 20$~s, 
at 0814~20:21   $260 \pm 25$~s. 
In general,  the rise time of the flares is in the range 220 - 440~s. 

We calculate the peak flux of the fireball as 
\begin{equation}
F_{pk}=F_{max}-F_{st},
\label{eq.Fpk} 
\end{equation}
where $F_{max}$ is the maximum flux of the star 
during the flare  and $F_{st}$ is the flux at the start of the flare.
We calculate  $F_{max}$ and $F_{st}$  for each band, using 
Bessel (1979) calibration  for the fluxes of a zero magnitude star.


In order to estimate the actual flux of the fireball, the observed magnitudes must  be corrected for 
the effects of interstellar extinction. 
The extinction towards AE~Aqr is low. 
La Dous (1991) found an $E(B-V) \approx 0.0$ from the weakness of 2200 \AA \  absorption feature.
Bruch (1992)  assumed  interstellar extinction $E_{B-V} =0.05$. 
The IRSA Galactic Reddening and Extinction Calculator (NASA/IPAC) 
gives for the interstellar extinction toward AE~Aqr 
\hskip 0.1cm 
a value of $E_{B-V} =  0.0610 \pm 0.0015$ and $A_V = 0.1891 $ mag. 
We adopt $E_{B-V} =  0.05$  and an extinction law as given in Cardelli et al. (1989).

The distance to AE~Aqr is estimated  as  $d=102$~pc (Friedjung 1997).  
Following the latest  reduction of the astrometric data as produced by the Hipparcos mission 
(van Leeuwen 2007) the parallax is  $11.61 \pm 2.72$ mas,  
which corresponds to a distance of $d=86^{+26}_{-16}$ pc.
This value is almost equal to the value $d=82$~pc (equivalent to a parallax 12.2~mas) calculated  by van Paradijs et al. (1989) 
on the basis of V-band surface brightness and spectral type.

On the basis of the measured fluxes we calculate the dereddened colours of the fireballs at the peak  of the flare
and list them in Table~\ref{Tab.res}. The  colours are similar to those of 
the flickering light source  in other cataclysmic variables (see Bruch 1992).

\subsection{Model of a fireball}

Pearson, Horne \& Skidmore (2003, 2005)  derived  analytic expressions for the continuum light curves 
and spectra of flaring and flickering events that occur over a wide range of astrophysical systems. 
They compared these results to data taken from the cataclysmic
variables AE~Aqr, SS Cygni and also from SN 1987A, deriving physical parameters for the material involved. 
They  have shown that the observed flare spectrum 
and evolution of AE Aqr is reproducible with an isothermal fireball 
with Population II abundances. Here we transcribe only a part of their equations. The interested reader 
is referred to their papers for the detailed derivations.

The basic assumptions are:
 (i)  the flares of AE~Aqr are due to the appearance and expansion of blobs (fireballs);
 (ii) the blobs represent spherically symmetric expansion 
of a Gaussian density profile with radial velocity proportional to the distance from the center of the  expansion; 
 (iii) the fireballs are isothermal.

Following Pearson et al. (2005),  the dimensionless time  $\beta$ is defined as
\begin{equation}
  \beta = 1 + H(t-t_{pk}),
\end{equation}
where $t_{pk}$ is the time of the peak of the flare, $H$ is an  
\textquotedblleft expansion  constant\textquotedblright\   setting the speed of
the expansion. 
The dimensionless time  $\beta$ is also the expansion factor being  the constant of proportionality 
between the current and peak scale length $a_{pk}$:
\begin{equation}
  \beta \equiv  \frac{a}{a_{pk}}.
\end{equation}
The central  density of the fireball is 
\begin{equation}
  \rho = \frac{M}{ (\pi a^2)^{3/2}},
  \label{rho}
\end{equation}
where M is the total mass involved in the expansion (which we call fireball mass).
The speed of expansion at  $a$ is:
\begin{equation}
 v = H a.
 \label{vHa}  
\end{equation}
The optical depth  parallel to the observer's line of sight is
\begin{eqnarray}
\tau(y) & = & - \int_{\infty}^{-\infty} \kappa\ {\rm d}x \label{eqn:tauint}\\
& = & \left[\frac{\kappa_1\ \epsilon}{T^{\frac{1}{2}}\ \nu^{3}}
\frac{M^{2} }{\pi^{3}\ a^{5}}\right]\ e^{-2(\frac{y}{a})^{2}}
\int_{-\infty}^{\infty} e^{-2(\frac{x}{a})^{2}}
\ {\rm d}\left(\frac{x}{a}\right) \nonumber \\
& &
\label{eq.tauy}\\
&=& \tau_{0}\ e^{-2(\frac{y}{a})^{2}},  
\label{eqn:tauy2}
\end{eqnarray}
where $y$ is the impact parameter (the distance from the fireball center perpendicular to the line of sight), 
$\kappa_1$ is the linear absorption coefficient, 
$\epsilon$ is the correction for stimulated emission 
\begin{equation} 
\epsilon = 1 - e^{(h\nu / k T)} , 
\end{equation}
$T$  is the fireball temperature, $a$  is the length-scale (which we call fireball size).

The optical depth on the line of sight through the center of the fireball ($y=0$) is  
\begin{equation}
\tau_{0}=\frac{\kappa_1\  \epsilon\  M^{2}}{ 2^{1/2}\  T_0^{1/2} \ \nu^{3} \ \pi^{5/2} \ a^5 } .
\label{eq.tau0}
\end{equation}
The emission of the fireball is:
\begin{equation}
f_\nu=\frac{\pi\ a^{2}\ B_\nu(T)}{2\ d^{2}} \ S(\tau_0), 
\label{eqn.sum}
\end{equation}
where 
\begin{equation}
S(\tau_0) = \sum_{n=1}^{\infty} \frac{(-1)^{(n+1)}\ \tau_{0}^{n}} {n\ n!},
\label{eqn.satur}
\end{equation}
is  the "saturation function", plotted in Fig.~1 of Pearson et al. (2005).

\subsection{Parameters of the fireballs}

To evaluate  the fireball parameters $M, T, a, H, \rho$ and  $v$,  we performed the following:

1. We measure $F_{max}$, $F_{st}$, and calculate $F_{pk}$ (Eq.\ref{eq.Fpk}) in $UBVRI$ bands.
In Fig.\ref{fig.flux}, we plot the calculated peak fluxes,   
corrected for the interstellar extinction. 

2. Using a black body fit ($nfit1d$ routine of IRAF), we calculate the 
temperature of the fireball.

3. We calculate the size of the fireball at the peak, $a_{pk}$,
using Eq.\ref{eqn.sum} and  the peak optical depth $\tau_{0} = 6.8202$  (see Eq. 33 of Pearson et al. 2005). 

4. We estimate the mass of the fireball
using Eq.\ref{eq.tau0}. The estimates are done for 
fireballs with  Population II  abundances,  $\kappa_1 = 1.27 \times 10^{52}$ m$^{-1}$.

5. Fitting the observed light curves (see Fig.\ref{fig.models}),
we calculate  the expansion constant $H$.  

6. We calculate  the expansion velocity 
from Eq.\ref{vHa}  and  the central  density  from Eq.\ref{rho}.

The estimated parameters with the corresponding errors are given in Table~\ref{Tab.res}.
The observed V band time evolution of the fireballs together with 
our best fit light curves are plotted in Fig.\ref{fig.models}. As can be
seen in  Fig.\ref{fig.models}, we achieved good agreement with the observed V band light curves of the optical flares
using  fireballs  with  a temperature T $\sim 15000$ K and mass $M \sim 10^{20}$~g.

The effects of the  uncertainty of the distance to the values listed in Table~\ref{Tab.res} are:
a change of the distance from d=86~pc to d=102~pc 
leads to an increase of $a$ with 19\%, of $M$ -- with 53\%, of $v$ -- with 19\% ,
and a decrease of $\rho$ with 8\%.

\section{Discussion}

\subsection{Comparison with spectral data}

The previous investigations demonstrated that the colour indices of the flares in AE~Aqr 
are close to those of a black body with $T \sim 15000 - 20000$ K (Beskrovnaya et al. 1996). 
LTE fits to the continuum of a large flare  yielded a temperature between 
$\sim 7 - 15 \times 10^3$~K (Welsh,  Horne \& Oke  1993). 

Skidmore et al. (2003) have obtained 
high quality spectroscopic data on  the 10~m  Keck telescope for one small flare.
They extracted and analyzed the spectrum 
of this flare, which   was  later modeled by Pearson et al. (2003) to obtain the 
fireball parameters. 
They used a distance of 100~pc. Their parameters recalculated for d=86~pc 
are  M=$1\times 10^{20}$~g, $a_0= 8.1 \times 10^9$ cm,  $\rho_0 = 3.5 \times 10^{-11}$ g~cm$^{-3}$,
and $v(a_0) = 144$ \kms.
The temperature  and expansion constant remain unchanged. 

Here we give a comparison of our  (summarized in Table~\ref{Tab.res}) with their results:

- temperature:  their estimate T=18000~K is well within our range of temperatures   10000 - 25000 K; 

- size: their value is bigger but similar  to our range $(3-7) \times 10^9$~cm.

- mass: their estimate  is in the lower part of our range  $(7 - 90) \times 10^{19}$~g.

- expansion velocity:   their value is in the middle of our range  100 - 350 \kms . 

- expansion constant:   their value  1/H = 560~s, is higher but similar to our range 210 - 375 s. 

- density:  their value  is lower than our range $(4 - 7) \times 10^{-10}$~g~cm$^{-3}$. 

We have good agreement with the spectroscopic results of Pearson et al. (2003). 
As expected  their  mass  and density are lower 
because they  have modeled  one smaller flare with amplitude $\sim 0.1$~mag, 
while our flares have  larger amplitudes ($\Delta V > 0.2$ mag).


\subsection{Mass transfer rate and fireball mass}

An estimation of the expected fireball mass can be obtained  from the mass accretion rate
and typical time between the flares.

Pearson et al. (2003) have estimated that the mass donor in AE~Aqr transfers  mass at a 
rate $\dot M = 3 \times 10^{17}$ g s$^{-1}$, using the standard evolutionary equation.
From a comparison of simulated and observed H$\alpha$ Doppler tomograms, 
Ikhsanov et  al. (2004) provided the mean value of the mass-transfer rate 
$\dot M = 5 \times 10^{16}$ g s$^{-1}$.
For total 18h42min  of observations, we detect 14 fireballs in the optical bands
which means about  0.75  fireball per hour and expected typical mass of 
the fireball $M \approx (1-20) \times 10^{20}$ g. 

We calculate the mass of the fireballs in the range $(0.7 - 9.0) \times 10^{20}$~g (see Table~\ref{Tab.res}), 
which agree with the above estimation obtained on the basis of the mass transfer rate. 
We measure the parameters of the fireballs that produce the most prominent peaks in the optical bands. 
These are probably the most massive blobs. Smaller blobs producing smaller flares should also exist. 

\subsection{Fireball vs. magnetosphere size}

The mass of the white dwarf in AE~Aqr is estimated as $M_{WD} = 0.63$~\msun.  
\hskip 0.8mm 
Following  mass-radius relation for magnetic white dwarfs  (Suh \& Mathews 2000)  
its radius  should be  $R_{WD} = 8.3 \times 10^8$~cm.
It means that the calculated fireball peak size is  $\sim 4 - 8 $~$R_{WD}$.


Choi \& Yi (2000), estimated  the magnetic field of the white dwarf $3\times 10^5$~G
on  basis of the quiescent X-ray and UV emission. It could be as high as $5 \times 10^7$~G
(Ikhsanov et al. 2004 and references therein).  
Using the range of parameters for the magnetic field and mass transfer rate
we calculate  the radius of the magnetosphere is \simlt$20 \; R_{WD}$, 
which means that the typical blobs have size less than 40\% of the magnetosphere radius.


\subsection{Blobs ejected from white dwarfs}

There are two other white dwarfs in longer period binary systems, for which blobs are observed and 
a propeller acting white dwarf is supposed.

{\bf RS~Oph:} 
\hskip 0.5cm 
 High-velocity emission components are observed in
the Balmer line wings in the recurrent nova 
RS~Oph (Iijima et al. 1994). The emission components most
probably originate in high velocity blobs of matter ejected  via a jet mechanism or centrifugally 
expelled by the white dwarf magnetosphere (Zajczyk et al. 2008). 
The  RS~Oph  blobs  have  mass  $M \approx 2 \times 10^{-9}$  \msun\ $= 4 \times 10^{24}$ g, velocity range
550 - 2060 \kms, and  $\sim \, 1$ month ejection interval (Zamanov et al. 2005). 
This  means similar ejection velocity,  $1000$ times higher mass 
and  700 times longer ejection interval than the AE~Aqr blobs.

{\bf CH~Cyg:} Similar blobs have been observed in 1994 in CH~Cyg.  
The spectroscopy of CH~Cyg with time resolution $10$ min, 
reported by Tomov et al. (1996)  shows that
emission components appeared and disappeared
on both sides of the $H\beta$ emission. 
They suggested that these emission components 
originate in small, short-lived blobs of material ejected in different directions, 
due to a magnetic propeller state. 
The blobs have  typical ejection timescale  3-5 hours (0.2-0.3 blobs per hour) and
velocity  up to $2000$~\kms. We estimate the mass of the blobs 

      $M_{\rm CHCyg} \approx  0.018 \; M_{\rm RSOph}  =  7 \times 10^{22}$ g. \\
This  means  70 times higher mass, 
and  $\sim 3$ times longer ejection interval than the AE~Aqr blobs.

It is worth noting that  
\hskip 0.4mm
  {\bf (i)} the mass accretion rate in CH~Cyg is similar 
or higher than in AE~Aqr - $3 \times 10^{16} - 10^{18}$~g~s$^{-1}$
 (Mikolajewska et al. 1988;  Wheatley \& Kallman 2006),  
 while in  RS~Oph it is about 100 times higher ($\sim 2 \times 10^{19}$~g~s$^{-1}$,  
 Osborne et al. 2011); 
 {\bf (ii)} the  white dwarfs in symbiotic stars probably  rotate much slower
 ($\sim 100$ times), having spin period of the order of 1 hour  
 (Sokoloski \& Bildsten 1999,  Formiggini \& Leibowitz 2009).

We pose the question: are there different mechanisms for a white dwarf to eject blobs
\hskip 0.1cm 
or it 
\hskip 0.1cm 
 is the same phenomenon (magnetic propeller), just  scaled with the spin period and mass accretion rate?

\section{Conclusions}

Using 4 telescopes, we performed simultaneous  observations in  5 bands $(UBVRI)$ 
of the flare activity  of  the cataclysmic variable AE Aqr.  
The main results of our work are: 

(1) For a total of  18.7 hours of observations we detected 14 flares. 
The rise time of the flares is in the range  220-440~s.

(2) For 5 individual fireballs, we calculated the peak fluxes in $(UBVRI)$   bands, rise times,  colours, 
and modeled their evolution. 
At the peaks of the optical emission, the dereddened colours of the fireballs are:
$(U-B)_0$ in the range 0.8-1.4, $(B-V)_0 \sim $    0.03-0.24.  

(3) Adopting  the model of an isothermically expanding  ball of gas, 
we find for the individual fireballs a temperature in the range 10000 -- 25000~K, 
mass  7-90$\times 10^{19}$~g,  size 3-7$\times 10^9$~cm,  
constant of expansion  1/H = 210 - 370~s$^{-1}$
(using a distance to AE~Aqr of $d=86$~pc and  interstellar extinction $E_{B-V} =0.05$). 
These values refer to the peak of the flares observed in the optical bands.

We also briefly discuss  the possible relation of the blobs of AE~Aqr with those detected in two symbiotic stars.

\section*{Acknowledgments}

We acknowledge the partial support by Bulgarian National Science Fund (DO~02-85 and DMU 03/131). 

We thank M.F. Bode  for an initial reading of the draft manuscript and the referee for useful comments.

This research has made use of the NASA/IPAC Infrared Science Archive, which is operated by 
the Jet Propulsion Laboratory, California Institute of Technology, under contract with 
the National Aeronautics and Space Administration.


\begin{thebibliography}{}

\bibitem[Bastian et al.(1988)]{1988ApJ...324..431B} Bastian, T.~S., Dulk, G.~A., \& Chanmugam, G.\ 1988, ApJ, 324, 431 

\bibitem[Beskrovnaya et al.(1996)]{1996A&A...307..840B} Beskrovnaya, N.~G., Ikhsanov, N.~R., 
Bruch, A., \& Shakhovskoy, N.~M.\ 1996, A\&A, 307, 840 

\bibitem[Bessell(1979)]{1979PASP...91..589B} Bessell, M.~S.\ 1979, PASP, 91, 589 

\bibitem[Bookbinder \& Lamb(1987)]{1987ApJ...323L.131B} Bookbinder, J.~A., \& Lamb, D.~Q.\ 1987, ApJ, 323, L131 

\bibitem[Bowden et al.(1992)]{1992APh.....1...47B} Bowden, C.~C.~G., Bradbury, S.~M., 
Chadwick, P.~M., et al.\ 1992, Astroparticle Physics, 1,  47 


\bibitem[Bruch(1992)]{1992A&A...266..237B} Bruch, A.\ 1992, A\&A, 266, 237 


\bibitem[Bruch \& Grutter(1997)]{1997AcA....47..307B} Bruch, A., \& Grutter, M.\ 1997, AcA, 47, 307 

\bibitem[Cardelli et al.(1989)]{1989ApJ...345..245C} Cardelli, J.~A., Clayton, G.~C., \& Mathis, J.~S.\ 1989, ApJ, 345, 245 

\bibitem[Casares et al.(1996)]{1996MNRAS.282..182C} Casares, J., Mouchet,  M., Martinez-Pais, I.~G., 
\& Harlaftis, E.~T.\ 1996, MNRAS, 282, 182 

\bibitem[Chincarini \& Walker(1981)]{1981A&A...104...24C} Chincarini, G., \& Walker, M.~F.\ 1981, A\&A, 104, 24 

\bibitem[Choi \& Yi(2000)]{2000ApJ...538..862C} Choi, C.-S., \& Yi, I.\ 2000, ApJ, 538, 862 


\bibitem[de Jager et al.(1994)]{1994MNRAS.267..577D} de Jager, O.~C., 
Meintjes, P.~J., O'Donoghue, D., \& Robinson, E.~L.\ 1994, MNRAS, 267, 577 

\bibitem[Echevarr{\'{\i}}a et al.(2008)]{2008MNRAS.387.1563E} Echevarr{\'{\i}}a, J., Smith, R.~C., Costero, R., Zharikov, S., 
\& Michel, R.\ 2008, MNRAS, 387, 1563 


\bibitem[Formiggini \& Leibowitz(2009)]{2009MNRAS.396.1507F} Formiggini, L., \& Leibowitz, E.~M.\ 2009, MNRAS, 396, 1507 

\bibitem[Friedjung(1997)]{1997NewA....2..319F} Friedjung, M.\ 1997, New Astronomy, 2, 319 


\bibitem[Henize(1949)]{1949AJ.....54...89H} Henize, K.~G.\ 1949, AJ, 54, 89 

\bibitem[Iijima et al.(1994)]{1994A&A...283..919I} Iijima, T., Strafella, F., Sabbadin, F., \& Bianchini, A.\ 1994,  A\&A, 283, 919 

\bibitem[Ikhsanov et al.(2004)]{2004A&A...421.1131I} Ikhsanov, N.~R., Neustroev, V.~V., \& Beskrovnaya, N.~G.\ 2004, A\&A, 421, 1131 

\bibitem[Jameson et al.(1980)]{1980MNRAS.191..559J} Jameson, R.~F., King, A.~R., \& Sherrington, M.~R.\ 1980, MNRAS, 191, 559 

\bibitem[La Dous(1991)]{1991A&A...252..100L} La Dous, C.\ 1991, A\&A, 252, 100 

\bibitem[Mauche et al.(1997)]{1997ApJ...477..832M} Mauche, C.~W., Lee, Y.~P., \& Kallman, T.~R.\ 1997, ApJ, 477, 832 

\bibitem[Mauche et al.(2011)]{2011arXiv1111.1190M} Mauche, C.~W., Abada-Simon, M., Desmurs, J.-F., et al., 
\ 2011, arXiv:1111.1190,  (Mem.  SAIt. 83, in press)  

\bibitem[Meintjes et al.(1992)]{1992ApJ...401..325M} Meintjes, P.~J., Raubenheimer, B.~C., de Jager, O.~C., et al.\ 1992, ApJ, 401, 325 


\bibitem[Mikolajewska et al.(1988)]{1988A&A...198..150M} Mikolajewska, J., Selvelli, P.~L., \& Hack, M.\ 1988, A\&A, 198, 150 


\bibitem[Oruru \& Meintjes(2012)]{2012MNRAS.421.1557O} Oruru, B., \& Meintjes, P.~J.\ 2012, MNRAS, 421, 1557 

\bibitem[Oruru \& Meintjes(2011)]{2011fxts.confE..63O} Oruru, B., \& Meintjes, P.~J.\ 2011, 
Fast X-ray Timing and Spectroscopy at Extreme Count Rates (HTRS 2011),  

\bibitem[Osborne et al.(2011)]{2011ApJ...727..124O} Osborne, J.~P., Page, K.~L., Beardmore, A.~P., Bode, M. F., 
Goad, M. R., O'Brien, T. J.,  Starrfield, S., Rauch, T., et al.\ 2011, ApJ, 727, 124 

\bibitem[Patterson(1979)]{1979ApJ...234..978P} Patterson, J.\ 1979, ApJ, 234, 978 

\bibitem[Pearson et al.(2003)]{2003MNRAS.338.1067P} Pearson, K.~J., Horne, K., \& Skidmore, W.\ 2003, MNRAS, 338, 1067 


\bibitem[Pearson et al.(2005)]{2005ApJ...619..999P} Pearson, K.~J., Horne, K., \& Skidmore, W.\ 2005, ApJ, 619, 999 

\bibitem[Schenker et al.(2002)]{2002MNRAS.337.1105S} Schenker, K., King, A.~R., Kolb, U., 
Wynn, G.~A., \& Zhang, Z.\ 2002, MNRAS, 337, 1105 

\bibitem[Skidmore et al.(2003)]{2003MNRAS.338.1057S} Skidmore, W., O'Brien, 
K., Horne, K., Gomer, R., Oke, J. B.,  Pearson, K. J.  \ 2003, MNRAS, 338, 1057 

\bibitem[Sokoloski \& Bildsten(1999)]{1999ApJ...517..919S} Sokoloski, J.~L., \& Bildsten, L.\ 1999, ApJ, 517, 919 

\bibitem[Suh \& Mathews(2000)]{2000ApJ...530..949S} Suh, I.-S., \& Mathews, G.~J.\ 2000, ApJ, 530, 949 


\bibitem[Tomov et al.(1996)]{1996MNRAS.278..542T} Tomov, T., Kolev, D., Munari, U., \& Antov, A.\ 1996, MNRAS, 278, 542 

\bibitem[van Leeuwen(2007)]{2007A&A...474..653V} van Leeuwen, F.\ 2007, A\& A, 474, 653 

\bibitem[van Paradijs et al.(1989)]{1989A&AS...79..205V} van Paradijs, J., van Amerongen, S., \& Kraakman, H.\ 1989, A\&AS, 79, 205 

\bibitem[Welsh et al.(1993)]{1993ApJ...406..229W} Welsh, W.~F., Horne, K., \& Oke, J.~B.\ 1993, ApJ, 406, 229

\bibitem[Wheatley \& Kallman(2006)]{2006MNRAS.372.1602W} Wheatley, P.~J., \& Kallman, T.~R.\ 2006, MNRAS, 372, 1602 

\bibitem[Wynn et al.(1997)]{1997MNRAS.286..436W} Wynn, G.~A., King, A.~R., 
\& Horne, K.\ 1997, MNRAS, 286, 436 

\bibitem[Zajczyk et al.(2008)]{2008ASPC..401..106Z} Zajczyk, A., Tomov, T., 
Mikolajewski, M.,  Bull, Ch., Kolev, D., Cikala, M., Georgiev, L., Galazufdinov, G.,\ 
2008, RS Ophiuchi (2006) and the Recurrent Nova 
Phenomenon, 401, 106 

\bibitem[Zamanov et al.(2005)]{2005MNRAS.363L..26Z} Zamanov, R.~K., Bode, M.~F., Tomov, N.~A., \& Porter, J.~M.\ 2005, MNRAS, 363, L26 

\bibitem[Zinner(1938)]{1938AN....265..345Z} Zinner, E.\ 1938, Astronomische Nachrichten, 265, 345 


\end{thebibliography}
\end{document}